# Effects of spatial quantization and Rabi-shifted resonances in single and double excitation of quantum wells and wires induced by few-photon optical field

A.N. Vasil'ev[1,2] and O.V. Tikhonova[1,2]

[1] Faculty of Physics, Lomonosov Moscow State University, Moscow, Russia
[2] Skobeltsyn Institute of Nuclear Physics, Lomonosov Moscow State University, Moscow, Russia
e-mail: anvasiliev52@gmail.com

**Abstract** We develop a fully quantum theoretical approach which describes the dynamics of Frenkel excitons and bi-excitons induced by few photon quantum light in a quantum well or wire (atomic chain) of finite size. The eigenenergies and eigenfunctions of the coupled exciton-photon states in a multiatomic system are found and the role of spatial confinement as well as the energy quantization effects in 1D and 2D cases is analyzed. Due to the spatial quantization, the excitation process is found to consist in the Rabi-like oscillations between the collective symmetric states characterized by discrete energy levels and arising in the picture of the ladder bosonic operators. At the same time, the enhanced excitation of additional states with energy close to the upper polariton branch is revealed. The found new effect is referred to as the formation of Rabi-shifted resonances and is analyzed in details. Such states are shown to influence dramatically on the dynamics of excitation especially in the limit of large times.

## 1. Introduction

Interaction of light with semiconductor quantum systems is the basis of many practical applications and optoelectronic devices [1-3]. For the systems with reduced dimensionality the formation of excitons should be considered and the effects of spatial quantization are found to be of great importance [4]. During the interaction of a photon with a quantum well or a quantum wire its wave vector usually appears to be negligibly small in comparison with inverse size of the system. Therefore the long-wave approximation is valid [5-7] and the electron transitions from the valence band to the conduction band occur with conservation of momentum such as the exciton is formed with a total quasi momentum equal to zero (only "vertical" transitions in wave-vector space are allowed) [7-9]. For this reason, the creation of excitons in semiconductor systems is often treated in a ladder picture in terms of bosonic creation/annihilation operators [10-16]. However actually not the only state with zero quasi momentum but a set of states in an infinitely small vicinity around it are involved in the transitions. This fact leads to difficulties of theoretical treatment since strictly speaking an infinite number of states should be taken into account in this case.

In systems with a finite size $L$, the momentum conservation law is not strictly fulfilled, and the quasi momentum is efficiently conserved only up to the value of $1/L$. In quantum wells and wires with size in some directions significantly larger than atomic ones, but finite, the spatial quantization results in a discretization of exciton energy states that are distributed quasi-continuously with distance between individual levels decreasing with growing $L$. This peculiarity leads to specific features of the exciton dynamics which is still not analyzed and described correctly.

In this work, we investigate the interaction of a quantum photon field in a high-Q resonator with a quantum well or a quantum wire of finite size, depending on the detuning of the resonator with respect to the minimal exciton transition frequency. We consider the case of one or two photons in the initial field state of a single photon mode. In the case of very-high-Q resonator the interaction with single photon mode can be very efficient and corresponding Rabi period can be much shorter than relaxation times for electron system [8,17,18]. In this case one has to take into account the discrete structure of the exciton energy spectrum.

Starting from the microscopic analysis we develop a fully quantum theoretical approach which describes the field-induced dynamics of Frenkel excitons in a quantum well or wire (atomic chain) of finite size. We have found the energies and eigenfunctions of collective excitonic states induced by light in a multiatomic system and analyze the role of spatial confinement and energy quantization effects in 1D and 2D cases. The obtained results are compared with the description of excitons in terms of the ladder bosonic operators. We show that dynamics of the system demonstrates not only polariton-like behavior but also the enhanced excitation of energy states above the photon energy by the Rabi frequency takes place. This found new effect is referred to as the formation of Rabi-shifted resonances and is analyzed in details.

## 2. Theoretical approach

For simplicity, we consider the Frenkel exciton in a system of two-level atoms in a chain of length $N$ (in 1D case) or square lattice of $N \times N$ atoms (in 2D case) and take into account the dipole-dipole transfer of excitation between neighbor atoms.

In the case of atomic chain the Hamiltonian of the Frenkel exciton interacting with quantum field mode consists of the non-perturbed photon and exciton energies and the interaction term correspondently and is given by:

$$\hat{H} = \hbar\omega b^\dagger b + \left( \sum_{n=1}^{N} \varepsilon a_n^\dagger a_n - \sum_{n=1}^{N-1} w a_n^\dagger a_{n+1} - \sum_{n=2}^{N} w a_n^\dagger a_{n-1} \right) + g\sum_{n=1}^{N}\left( b a_n^\dagger + b^\dagger a_n \right) \equiv \hat{H}_{ph} + \hat{H}_0 + \hat{H}_{int}, \quad (1)$$

where $b^\dagger(b)$ stand for bosonic creation (annihilation) photon operators in a resonator mode, $a_n^\dagger(a_n)$ are the fermionic operators of creation (annihilation) of excitation of the atom with number "$n$" with double excitation of the same atom being forbidden. The parameter $w$ characterizes the strength of dipole-dipole excitation transfer which depends on the field polarization direction in relation to the chain orientation, $\varepsilon$ and $\hbar\omega$ stand for energies of atomic excitation and a photon respectively, and $g$ is the coupling constant of photon-atom interaction.

The wave functions $|1,k\rangle$ of a single exciton in the chain describing the case when one of the atoms is excited with probability amplitude $\chi_{nk}$ can be given by:

$$|1,k\rangle = \sum_{n=1}^{N}\chi_{nk}^{N} a_n^\dagger |0\rangle, \quad \chi_m^N \equiv \sqrt{\frac{2}{N+1}}\sin\left(\frac{\pi m}{N+1}\right). \quad (2)$$

Here $\chi_{nk}$ can be referred to the shape of the exciton mode and $k = 1,\ldots,N$ is the number of this mode. The obtained field-free solution can be easily found by periodic extension of the chain with inserting the $(N+1)$-th atom, which cannot receive any excitation. In contrast to [10-16], the found states (2) can't be described by bosonic ladder operators. The energy of this state can be found as follows:

$$E_{1,k} = \varepsilon - 2w\cos\left(\frac{\pi k}{N+1}\right), \quad (3)$$

and for large $N$ turns into the usually used parabolic dependence [9] on the quasi momentum $q = \pm \dfrac{\pi k}{d(N+1)}$ around the extremum $\varepsilon_0 = \varepsilon - 2w$ in the following form:

$$E_{1,k} \approx \varepsilon_0 + w\left(\dfrac{\pi k}{N+1}\right)^2 = \varepsilon_0 + \dfrac{\hbar^2 q^2}{2m_{eff}}, \qquad (4)$$

Here the effective mass is given by $m_{eff} = \hbar^2/(2wd^2)$, $d$ is the distance between atoms. For our system of finite size, the minimal exciton energy slightly exceeds $\varepsilon_0$ due to quantum-confinement effects and is equal to $\varepsilon_0 + w\pi^2/(N+1)^2$.

States with several excitons are obtained in a similar way, taking into account the fact that two excitations cannot exist on one atom. Thus, the energy of the biexciton state is given by:

$$E_{2,k_1,k_2} = 2\varepsilon - 2w\left(\cos\left(\dfrac{\pi k_1}{N+1}\right) + \cos\left(\dfrac{\pi k_2}{N+1}\right)\right), \quad k_1 = 1,\ldots N-1, \quad k_2 = k_1+1,\ldots N. \qquad (5)$$

It is assumed that $k_2 > k_1$ due to the indistinguishability of excitons. The total number of biexciton states is equal to $\dfrac{N!}{2!(N-2)!} = \dfrac{N(N-1)}{2}$, and their wave functions can be found in the form:

$$|2,k_1,k_2\rangle = \sum_{n_1=1}^{N-1}\sum_{n_2=n_1+1}^{N} \left(\chi_{n_1 k_1}^N \chi_{n_2 k_2}^N - \chi_{n_1 k_2}^N \chi_{n_2 k_1}^N\right) a_{n_1}^\dagger a_{n_2}^\dagger |0\rangle, \quad k_1 = 1,\ldots N-1, \quad k_2 = k_1+1,\ldots N. \qquad (6)$$

Again, due to the indistinguishability of excitons, the summation is carried out for $n_2 > n_1$.

The generalization to the 2D, 3D and $m$ exciton case is described in Supplementary material.

It can be easily seen that the operator of this interaction can be rewritten in the following form:

$$\hat{H}_{int} = g\sum_{n=1}^{N}\left(ba_n^\dagger + b^\dagger a_n\right) \equiv g\left(bR^\dagger + b^\dagger R\right), \qquad (7)$$

where the operator $R^\dagger = \sum_{n=1}^{N} a_n^\dagger$ is responsible for the creation of the excitation of one of the atoms in the chain with equal probability. It should be emphasized that this operator appears to be bosonic one and leads to the excitation of the symmetric "collective" excitonic mode in the chain $\dfrac{1}{\sqrt{N}} R^+ |0\rangle$ with zero value of quasi momentum of the exciton. At the same time the biexcitonic symmetric "collective" state can be found as $\dfrac{1}{\sqrt{2N(N-1)}} R^{+2}|0\rangle$. For this reason, the exciton dynamics in semiconductor is usually described in terms of the bosonic ladder operators and collective excitonic modes.

The interaction between the atomic chain and quantum field mode is presented in (1) in the dipole approximation using the rotative wave approximation (terms with fast optical oscillations are excluded). The selection rules of field-induced transitions lead to excitation of states only with odd quantum numbers $k = 1, 3,\ldots, 2\left\lfloor\dfrac{N-1}{2}\right\rfloor + 1$. Such states can be called "active". The oscillator strength for the transition to these states can be calculated analytically:

$$f_k = |\gamma_k|^2, \gamma_k = \langle 0|bR^\dagger + b^\dagger R|1,k\rangle = \sum_{n=1}^{N}\chi_{nk}^N = \frac{\sqrt{1-(-1)^k}}{\sqrt{N+1}}\cot\left(\frac{\pi k}{2(N+1)}\right). \quad (8)$$

It is important, that the oscillator strengths satisfy the sum rule $\sum_{k=1}^{N} f_k = N$. In case of $N \gg 1$, the dipole matrix element can be shown to be inversely proportional to the quantum number $k$ of excitonic mode (See Supplem. S1).

It should be emphasized that the superposition of free exciton eigenstates $|1,k\rangle$ with amplitudes equal to the transition matrix elements (8) explicitly gives collective symmetric excitonic state:

$$\sum_{\substack{k=1\\ \text{odd } k}}^{N} \gamma_k |1,k\rangle = \frac{1}{\sqrt{N}} R^\dagger |0\rangle. \quad (9)$$

Thus, a direct correspondence between transitions to a set of free excitonic states and the excitation of a collective symmetric mode is found. Does it mean that the formation and evolution of the excitation process in these two representations are equivalent? Below we analyze the role of collective states in excitation dynamics.

## 3. Results and discussion

In this section, we describe the dynamics of excitons induced in quantum wire and quantum well by few-photon quantum light in the case of the limited linear dimensions of the system (10-100 μ). The revealed dynamics is compared with the results obtained in the frame of bosonic ladder creation/annihilation operators of excitation of symmetric collective states. The spatial confinement of the system is found to induce new physical features of the time-dependent exciton behavior arising due to the formation of Rabi-shifted resonances and free exciton states.

### 3.1. Exact quasi-energy solution and dynamics of excitation

Let us first consider a 1D atomic chain and only one photon in the quantum field mode. The interaction is supposed to be switched on at the time moment $t=0$. The initial state of the system corresponds to first Fock state for the quantum field and zero number of excitons in the chain and is given by $|1;0\rangle = b^\dagger|0\rangle$. In this case, only "active" states describing one exciton and zero number of photons in the system $|0;1,k\rangle = |1,k\rangle$ (with odd $k$) are involved in the excitation process. The diagonalization of the Hamiltonian in terms of this basis leads to the zero value of the following determinant:

$$\begin{vmatrix} \hbar\omega - \lambda & g\gamma_1 & \cdots & g\gamma_M \\ g\gamma_1 & \varepsilon_1 - \lambda & 0 & 0 \\ \vdots & 0 & \ddots & 0 \\ g\gamma_M & 0 & 0 & \varepsilon_M - \lambda \end{vmatrix} = 0, \quad (10)$$

where $M$ is the total number of "active" states. In the one-dimensional case, $M = \lfloor N+1/2 \rfloor$ $j = 1\cdots M$. The secular equation which determines the quasi-energies of quantum polariton or the energies of exciton dressed by one-photon quantum field, can be written in the form:

$$(\hbar\omega - \lambda) = g^2 \sum_{j=1}^{M} \frac{\gamma_j^2}{\varepsilon_j - \lambda} = g^2 G(\lambda). \tag{11}$$

In the limit of large number of atoms the polariton energies can be found analytically (See Suppl..)

The eigenfunctions corresponding to the found from (11) quasi-energies $\lambda_m$ are given by:

$$\psi_m = P_m^{-1/2} \left\{ 1, \cdots, g \frac{\gamma_j}{(\varepsilon_j - \lambda_m)}, \cdots \right\}^T \tag{12}$$

with normalization factor $P_m$ given by:

$$P_m \equiv 1 + g^2 \sum_{j=1}^{M} \frac{\gamma_j^2}{(\varepsilon_j - \lambda_m)^2}. \tag{13}$$

The concept of the found "dressed" states appears to be very fruitful and its advantage comes from independence of such states since there are no transitions between them and only their "free" evolution takes place.

Figure 1a shows the spectrum of the found quasi-energies in dependence on detuning $\hbar\omega - \varepsilon_0$ with several important features being demonstrated. Instead of Rabi-originated two polariton dispersion curves well-known for two resonantly coupled states, there is only one lower polariton branch (curve 3) corresponding to the discrete quasi-energy $\lambda_0 < \varepsilon_0$ but a lot of the quasi-continuum energy levels with $\lambda_m > \varepsilon_0$ (above line 1) are found for each detuning. Red dashed line (2) describes photon state for zero $g$ (without interaction with chain, $\varepsilon = \hbar\omega$). Generally, all the quasi-energy states contribute to the time-dependent solution and the dynamics of the system is determined by the found quasi-energies $\lambda_m$ and correspondent wave-functions:

$$\varphi(t) = \sum_{m=0}^{M} b_m \psi_m \exp(i \lambda_m t / \hbar). \tag{14}$$

However, the weights of the quasi-energy states $b_m^2$ can be significantly different and are strongly determined by the initial wave function of the system.

It can be easily seen from (14) that these coefficients are determined by the normalization factors $P_m$:

$$b_m = \langle \psi_m | 1; 0 \rangle = P_m^{-1/2}. \tag{15}$$

These projections can be analytically estimated in the limit of large number of atoms (Suppl. 1).

The weights $b_m^2 = P_m^{-1}$ of different quasi-energy levels with $\lambda_m > \varepsilon_0$ involved in the solution for the chosen initial condition are shown on the inset in Fig. 1a by changing color intensity of red dots representing different levels with $\lambda_m$. As a result, the dots with maximum of $P_m^{-1/2}$ for positive $m$ and different detunings form the upper polariton branch shown by curve 4 in the main part of Fig 1a. The structure of this branch is more complex than that for lower polariton curve, due to its quasi-discrete character and avoid-crossing features arising from the influence of many close neighboring quasienergy states. All other eigenstates have negligibly small contribution of photon state and looks like pure exciton states.

Figure 1b shows more precisely the weights $b_m^2 = P_m^{-1}$ of quasi-energies for four values of detuning. One low quasienergy state with $\lambda_0 < \varepsilon_0$ and a set of states around some most pronounced quasi-energy $\lambda^* > \varepsilon_0$ are seen to contribute to the final solution. These additional quasi-energies around $\lambda^*$ appear in the solution due to arising resonances between a "dressed" state with eigenenergy $\lambda_m$ and free exciton level with a certain energy $\varepsilon_j$. These possible resonances between the free exciton energies and the upper polariton branch appear to be the key point of further discussed physical effects and are shown to change significantly the dynamics of the excitation process.

It should be emphasized that the two predominate quasienergies $\lambda_0$ and $\lambda^*$ are found to coincide exactly with the well-known lower and upper polariton energies arising in the two-level system and providing the Rabi oscillations between the initial state and collective symmetric mode $\frac{1}{\sqrt{N}}R^+|0>$. The correspondent weights of these "dressed states" are equal to the values $P_0^{-1}$ and $\sum_{m=1}^{M}P_m^{-1}=1-P_0^{-1}$ respectively and are marked in Fig. 1b by large discs.

The coupling between the initial state and collective symmetric mode $\frac{1}{\sqrt{N}}R^+|0>$ can be easily proved by the structure of lower polariton state with $\lambda_0$ found from (14) and given by:

$$C_0 = P_0^{-1/2}\left(|1;0\rangle + \frac{g\sqrt{N}}{\lambda_0-\varepsilon_0}\sum_{\substack{k=1\\ \text{odd }k}}^{N}\gamma_k|1,k\rangle\right) = P_0^{-1/2}\left(|1;0\rangle + \frac{g}{\lambda_0-\varepsilon_0}R^\dagger|0\rangle\right), \tag{16}$$

At the same time the quasi-energy state with $\lambda^*$ is not limited only to these two states since many neighboring quasienergy states are involved. For this reason at least couple of quasienergies around the peak value $\lambda^*$ should also be taken into account and are expected to contribute to the excitation dynamics.

As a result, except the found resonances the lower and upper polariton branches are responsible for the Rabi oscillations between the initial state and symmetric excitonic mode $\frac{1}{\sqrt{N}}R^+|0>$. Such temporal behavior is fully described in terms of the bosonic ladder operators. However, possible resonances between the free exciton energies and the upper polariton branch appear to change significantly the dynamics of the excitation process.

Let us discuss the dynamics of the system presented in Fig. 1c-h for three values of detuning. In the model of bosonic ladder operators the excitation induced by one photon is described as Rabi-oscillations between the ground state and collective symmetric state $\frac{1}{\sqrt{N}}R^+|0>$. Left column of Fig. 1 (c,e,g) shows the exact time-dependent population of excitons in the system (blue) and the difference between this population and the population of the collective symmetric state $\frac{1}{\sqrt{N}}R^+|0>$ (red). For more convenience, the abscissa has two time scales: times below 1.5 ns are shown in more details. At the initial stage of the excitation dynamics, the above-mentioned Rabi-like oscillations take place. However, for larger time the population of collective symmetric state $\frac{1}{\sqrt{N}}R^+|0>$ no longer fully describes the excitation process since some additional states are populated with increase of time (red line), and start to oscillate with frequency much lower than Rabi frequency. In a "dressed states" picture these are states with quasienergies neighboring to the $\lambda^*$ and characterized by highest weights (see Fig. 1b). In the free exciton basis, these oscillations are found to correspond to the population of non-perturbed excitonic "active" states around a certain number with energy close to upper polariton branch (see 3D plots in Fig. 1 d,f,h). The amplitude of the population of these resonance states increases when detuning becomes positive and decreases for negative detuning.

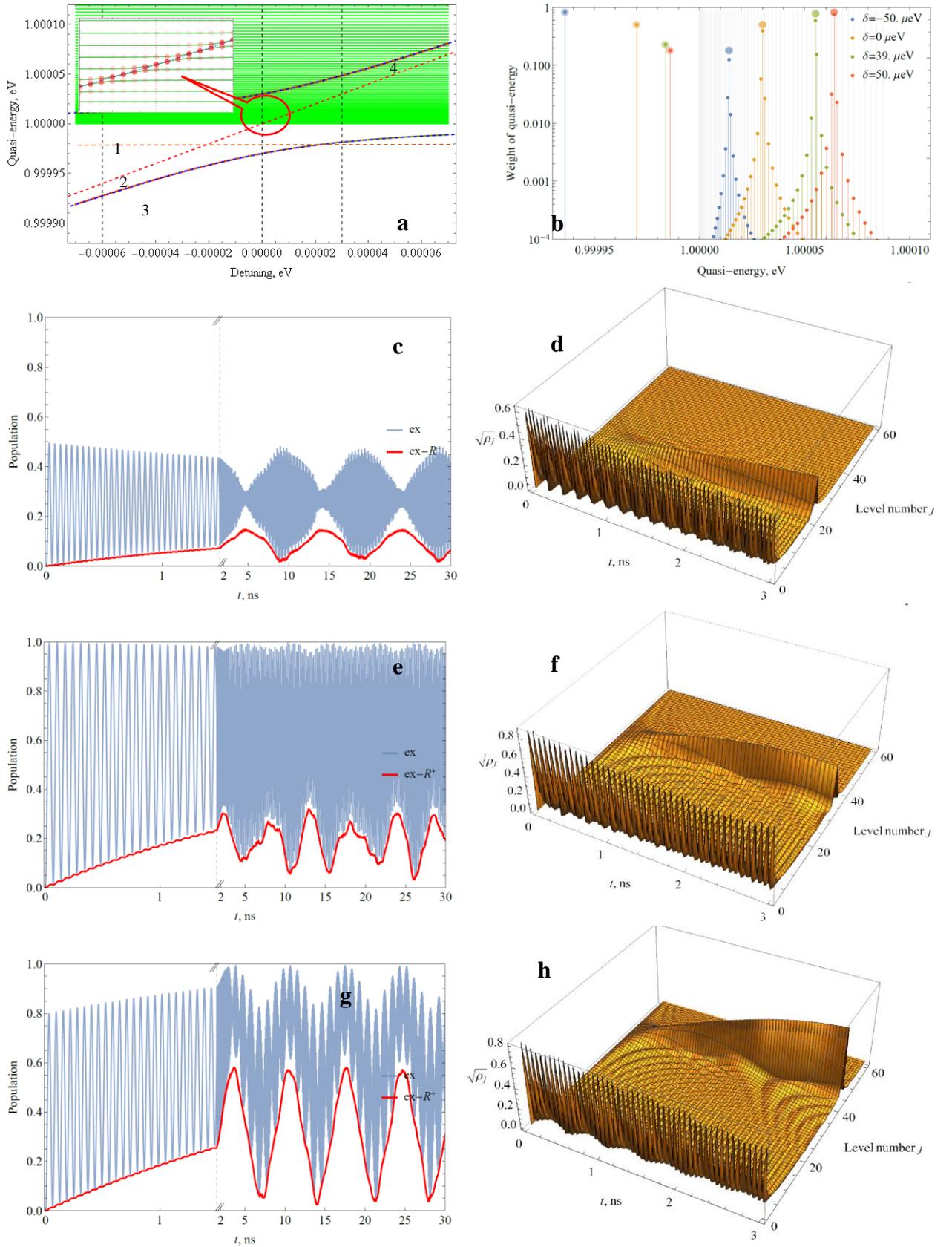

Figure 1. Simulation of 1D chain with $N = 20000$ atoms, $w = 0.25$ eV, $g\sqrt{N} = 30$ μeV. (**a**) Quasi-energies (green curves) depending on the detuning $\hbar\omega - \varepsilon_0$; the lower (curve 3) and upper (curve 4) polariton branches, $\varepsilon = \hbar\omega$ - red dashed line (2), $\varepsilon = \varepsilon_0$ – line (1). (**b**) Distribution of weights of quasi-energy states for different detuning values $\hbar\omega - \varepsilon_0$. Large circles correspond to positions and weights of quasi-energies for 2-level system (in $R^+|0\rangle$ approximation). (**c**) Time-

dependent population of excitons (blue) and the difference between this population and the population of the collective symmetric state $\frac{1}{\sqrt{N}}R^+|0>$ (red) calculated for negative detuning $\hbar\omega-\varepsilon_0 = -2g\sqrt{N}$ =-60 µeV. (**d**) Square root of population $|C_j(t)|$ of different excitonic states for the same parameters. (**e, f**) The same as (**c, d**) for zero detuning $\hbar\omega-\varepsilon_0 = 0$. (**g, h**) The same as (**c, d**) for positive detuning $\hbar\omega-\varepsilon_0 = g\sqrt{N} = 30$ µeV.

### 3.2. Comparison with 3-level system

To prove the idea about an important role of the found "resonance" states we consider a much simpler system characterized by two states resonantly coupled by photon with coupling constant $g$ (ground and main excited states, to be specific) and the upper state with detuning $\Delta$ above the main level with weaker interaction with light (coupling constant $g\mu$, $\mu \ll 1$). The scheme of considered energy levels is presented in Fig. 2a.

If we neglect the interaction with the upper level ($\mu = 0$), the system is characterized by two quasi-energies $\varepsilon_0 \pm g$ (see Fig. 2a, left panel). When we take into account the weak interaction with the upper level which is in resonance with the upper quasi energy $\varepsilon_0 + g$ ($\Delta = g$), this quasi-energy is split into two quasi-energies due to avoid-crossing effect (see Fig. 2a, right panel). The explicit values of quasienergies can be easily found in such a case and for $\mu \ll 1$ in linear approximation on $\mu$ appear to be equal to $\varepsilon_0 - g$, $\varepsilon_0 + g \pm g\mu/\sqrt{2}$. If at $t=0$ system is in the ground state and there is one photon in the field mode, the populations of ground $|C_0(t)|^2$ and main excited state $|C_1(t)|^2$ depends on time as follows:

$$|C_{0,1}(t)|^2 = \frac{3}{8} \pm \frac{1}{2}\cos(2\hbar^{-1}gt)\cos\left(\frac{\hbar^{-1}g\mu t}{\sqrt{2}}\right) + \frac{1}{8}\cos(\sqrt{2}\hbar^{-1}g\mu t), \qquad (19)$$

while the population of the upper Rabi-shifted resonance level $|C_2(t)|^2$ is given by:

$$|C_2(t)|^2 = \frac{1}{4}\left(1-\cos(\sqrt{2}\hbar^{-1}g\mu t)\right). \qquad (17)$$

The dynamics of these found populations is shown in Fig. 2b. The evolution of this system qualitatively represents the evolution of excited states in the chain (cf. Fig. 1). It is seen that initially weakly coupled upper level can efficiently exchange the energy with polariton formed by the main level and the photon, with maximum achievable population equal to one half. This polariton state is the analog of the polariton state in the 1D chain formed by coupling of collective symmetric state $\frac{1}{\sqrt{N}}R^+|0>$ and a photon, whereas the upper resonance level corresponds to a set of Rabi-shifted resonance levels in the chain. The number of quasi-energies involved in this resonance states is larger than in the case of three-level system, therefore the oscillations in the chain system are more irregular.

For the non-zero detuning $\delta = \hbar\omega-\varepsilon_0$ similar results can be obtained. The significant population of the upper state is achieved if this level is above the main one by $\Delta = \frac{1}{2}\left(\delta + \sqrt{4g^2+\delta^2}\right)$. The time-dependent population of the main and upper levels is presented in Fig. 2 c,d and an evident similarity between the obtained results and the exciton dynamics on Fig. 1 c,g can be found.

Thus, the coincidence (resonance) of free exciton energy with the upper polariton branch appears to perturb dramatically the quasienergy spectrum of the whole system and leads to significant population of these "resonance" states.

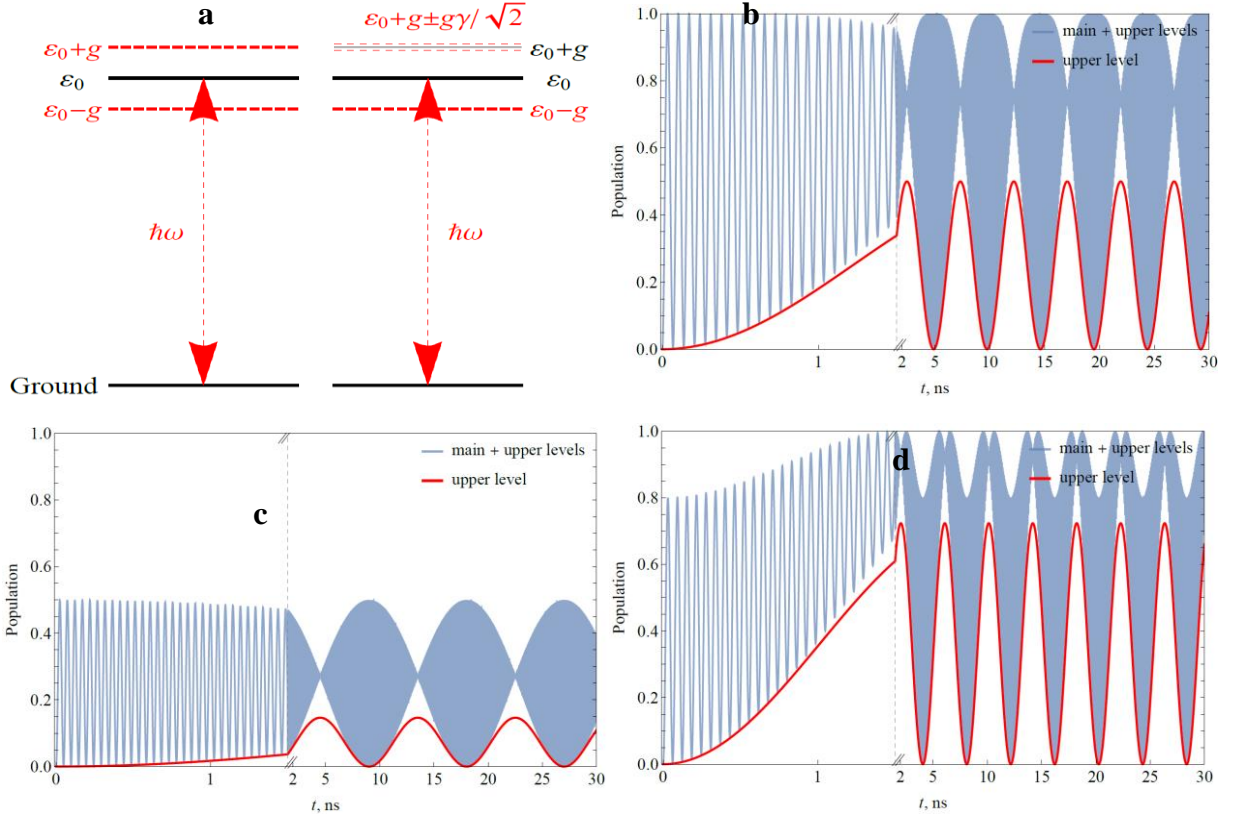

Figure 2. (**a**) Two-level (left) and three-level system (right) interacting with one photon. Red lines correspond to quasi-energies for coupling constants $g$ for main and $g\mu$ for upper Rabi-shifted resonance level. (**b-d**) Time-dependent population of the main level (blue) and upper level (red) for three-level system for $g=30$ µeV for zero detuning (**b**), $\delta = -2g$ (**c**) and $\delta = g$ (**d**).

### 3.3. Quantum well

Let us consider now a quantum well with monoatomic width. The formulas which describe fragment of crystal with simple cubic structure and $N_x \times N_y \times N_z$ atoms are presented in Supplementary Materials 2. In the case of significantly large transverse dimensions with $N_x \gg 1$ and $N_y \gg 1$ the field-free exciton states $|1, k_x, k_y>$ are characterized by two quantum numbers $k_x$ and $k_y$ with the correspondent energy written for large $N_x$, $N_y$ in the form:

$$E_{1,\{k_x k_y\}} = \varepsilon - 2w_x \cos\left(\frac{\pi k_x}{N_x+1}\right) - 2w_y \cos\left(\frac{\pi k_y}{N_y+1}\right) \approx$$
$$\approx \varepsilon - 2w_x - 2w_y + w_x\left(\frac{\pi k_x}{N_x+1}\right)^2 + w_y\left(\frac{\pi k_y}{N_y+1}\right)^2 \quad \text{for } k_x \ll N_x, k_y \ll N_y. \tag{18}$$

For simplicity we suppose that dipole moments of interacting atoms are directed perpendicular to the well. In this case the dipole-dipole transfer integrals are the same $w_x = w_y = w > 0$.

The dipole matrix element for field-induced transitions depends on both $k_x, k_y$ and is given by:

$$\gamma_{\{k_x k_y\}} = \frac{4\sqrt{\left(1-(-1)^{k_x}\right)\left(1-(-1)^{k_y}\right)(N_x+1)(N_y+1)}}{\pi^2 k_x k_y}. \tag{19}$$

The number of such "active" states approximately equals to $N_x N_y/4$, since $\gamma_{\{k_x k_y\}} \neq 0$ only for odd $k_x$ and $k_y$.

The theoretical analysis of the excitation in quantum well is rather similar to the case of the atomic chain using equation for quasienergies similar to Eq. (10). Unfortunately the sums in this equation cannot be evaluated analytically. Function $G(\lambda)$ can be evaluated using Kramers-Kronig relations from the absorption of the system proportional to $\sum_{\substack{k_x=1 \\ \text{odd}\, k_x}}^{N_x} \sum_{\substack{k_y=1 \\ \text{odd}\, k_y}}^{N_y} \gamma_{\{k_x k_y\}}^2 \delta\left(\varepsilon_{\{k_x k_y\}} - \hbar\omega\right)$. In case of quantum well main terms come from poles with $k_x=1$ or $k_y=1$. The total input from terms with $k_x \neq 1$ and $k_x \neq 1$ is about $1 - \frac{8}{\pi^2} \approx 0.19$. All poles are two-fold degenerated except one with $k_x = k_y = 1$.

Fig. 3 shows the dynamics of the excitations in case when only one photon is in the field mode and atoms in the well are unexcited at $t=0$. In order to compare the cases of the chain and quantum well, we should keep the total response of the chain ($g^2 N$) and the well ($g_w^2 N^2$) to be the same. In this case the branches of lower and upper polaritons obtained for the 2D well and 1D chain behave in a similar way (compare Fig. 3a and Fig. 1a), while the number of quasi-energy states is $N$ times higher in the case of the well, and the set of quasi-energies is much denser. The structure of weights of quasi-energies also appears to be similar for both cases, except the larger number of quasi-energies close to the upper polariton energy obtained for the quantum well (compare Fig. 3b and Fig. 1b). Some non-monotonous character of the weights of quasi-energies in a quantum well is connected with the mentioned above peculiarity of $\gamma_{\{k_x k_y\}}$ which have larger values for $k_x=1$ or $k_y=1$.

The excitation dynamics in a quantum well is found to be very similar to the case of 1D chain. It consists in the Rabi oscillations between the initial state and the collective mode

$$\frac{1}{\sqrt{N_x N_y}} R^\dagger |0\rangle = \frac{1}{\sqrt{N_x N_y}} \sum_{n_x=1}^{N_x} \sum_{n_y=1}^{N_y} a_{n_x n_y}^\dagger |0\rangle = \sum_{\substack{k_x=1 \\ \text{odd}\, k_x}}^{N_x} \sum_{\substack{k_y=1 \\ \text{odd}\, k_y}}^{N_y} \gamma_{\{k_x k_y\}}^2 \left|1, \{k_x k_y\}\right\rangle, \tag{20}$$

accompanied by the slow population of the exciton states which are in resonance with the upper polariton branch. These "resonant" states have almost the same energy proportional to $k_x^2 + k_y^2$ and the distribution of their population obtained at a certain instant of time is represented as a ring-shaped structure in 3D distributions of Fig. 3d,f,h with significant maxima pronounced at $k_x=1$ and $k_y=1$.

In contrast to 1D chain, in the case of quantum well there are several eigenstates with rather high weights around the upper polariton energy. Therefore the contribution of the resonance states to the exciton dynamics are mostly pronounced in the case of quantum well.

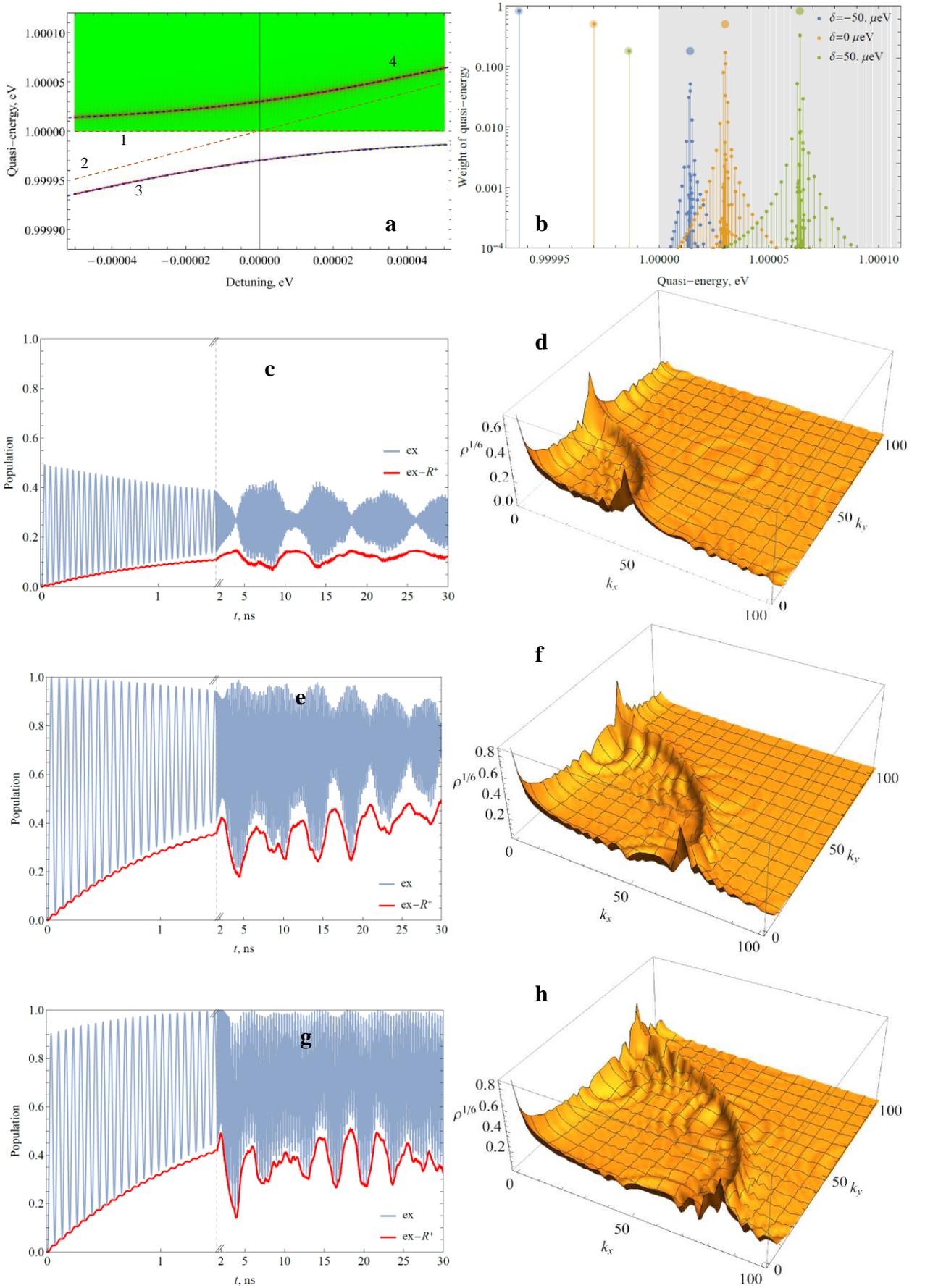

Figure 3. Simulation of 2D well with $N \times N = 20000 \times 20000$ atoms, $w = 0.25$ eV, $g_w N = 30$ μeV. (**a**) Quasi-energies (green curves) depending on the detuning $\hbar\omega - \varepsilon_0$. (**b**) Distribution of weights of quasi-energies for different detunings $\hbar\omega - \varepsilon_0$. Large circles correspond to positions and weights of quasi-energies for 2-level system ($R^+|0\rangle$ approximation). (**c**) Time-dependent population of excitons (blue) and the difference between this population and the population of the collective symmetric state $\frac{1}{\sqrt{N}}R^+|0>$ (red) calculated for negative detuning $\hbar\omega-\varepsilon_0 = -2g_wN = -60$ μeV. (**d**) Sixth root of population $|C_j(t)|^{1/3}$ of different excitonic states for the same parameters at $t = 0.6$ ns. (**e, f**) The same as (**c, d**) for zero detuning $\hbar\omega-\varepsilon_0 = 0$. (**g, h**) The same as (**c, d**) for positive detuning $\hbar\omega-\varepsilon_0 = g_wN = 30$ μeV.

## 4. Interaction of two photons with 1D chain

Let us consider the case when there are $m$ photons in the single photon mode at the zero time. In this case sequential excitation of the quantum wire is possible up to the production of $m$ excitons in the system. The detailed theoretical analysis of such process is given in the Suppl. 3. Hear we focus on the two photon initial state of the field mode providing exciton and bi-exciton production. For the chain with $N$ atoms there are $N$ excitonic and $N(N-1)/2$ bi-excitonic free states denoted $|1,k>$ and $|2,k_1,k_2>$ respectively. The matrix elements of photon–induced transitions between these states and found selection rules are discussed in Suppl. 3.

The spectrum of the "dressed" states and their weights found in the case of initially unexcited system are presented in Fig. 4ab respectively. The obtained polariton structure is found to consists now of 3 branches: lower (curve 4), intermediate (curve 5) and upper (curve 6) polariton branches which can be easily obtained in the frame of boson ladder operators and for zero detuning correspond to the quasienergies $\varepsilon_0 - 2g$, $\varepsilon_0$ and $\varepsilon_0 + 2g$ respectively. The correspondent collective symmetric modes are found as $\frac{1}{\sqrt{N}}R^+|0\rangle = \frac{1}{\sqrt{N}}\sum_{n=1}^{N}a_n^\dagger|0\rangle$ for excitonic state and $\sqrt{\frac{1}{N(N-1)}}(R^+)^2|0\rangle = \sqrt{\frac{2}{N(N-1)}}\sum_{n_1=1}^{N-1}\sum_{n_2=n_1+1}^{N}a_{n_1}^\dagger a_{n_2}^\dagger|0\rangle$ for bi-excitons. The structure of the found "dressed" states can be seen from the Fig. 4 d,f,h where the contribution of different free exciton and bi-exciton states is presented. The states with very small energies above $2\varepsilon_0$ are found to contribute strongly to the collective symmetric modes. For zero quasienergy state the contribution of exciton appears to be dramatically suppressed due to the destructive interference of photon-induced up and down-transitions. At the same time, the additional contribution of free exciton (blue points) and bi-exciton (yellow points) states with higher energies appears to be important which can be referred to as "resonant" states discussed in the previous Sections. The structure of these states is more complicated due to the parallel transitions from many exciton states to the same bi-exciton one and vice versa.

As a result, the time-dependent probability of excitons and bi-excitons in the system presented in Fig. 4 c, e ,g is mainly caused by dynamics of population of the collective symmetric modes as well as the involved "resonance" states. The population of these resonant states is generally higher than in the case of one photon in the initial state. However their role can be suppressed by growing the number of atoms, i. e. by passing to the limit of the infinitely long chain. It should be also noticed that the dynamics of the system excitation is revealed to be very sensitive to the field frequency and the detuning value. For negative detuning the contribution of "resonant" states appears to be rather small and the excitation process is correctly described in terms of the bosonic ladder operators. In this case significant population is found to remain in the initial unexcited state of the system. In contrast, for zero and positive detunings the system tends to be excited with the predominating contribution of bi-exciton production. The population of excitons tends to be suppressed due to the destructive interference of transitions involving these states.

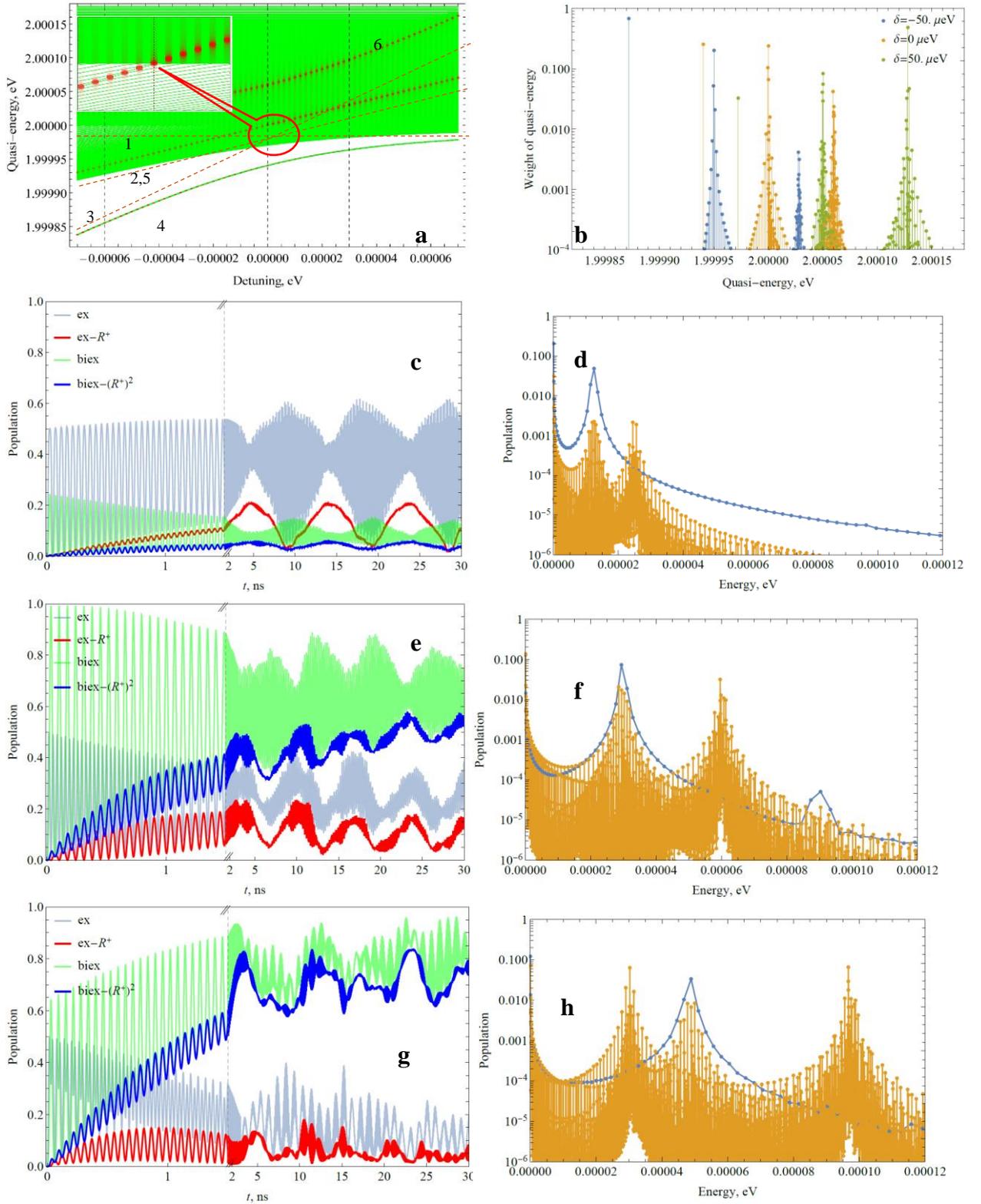

Figure 4. Simulation of 1D chain with $N = 20000$ atoms, $w = 0.25$ eV, $g\sqrt{N} = 30$ μeV for 2 photons in the initial state. (**a**) Quasi-energies (green curves) depending on the detuning $\hbar\omega - \varepsilon_0$. (**b**) Distribution of weights of quasi-energies for different detunings $\hbar\omega - \varepsilon_0$. (**c**) Time-dependent population of excitons (gray) and bi-excitons (green) as well as the difference between these populations and the population of correspondent collective symmetric states (red and violet respectively) obtained for negative detuning $\hbar\omega - \varepsilon_0 = -2g\sqrt{N} = -60$ μeV. (**d**) Population of different single excitonic states (blue) and two-excitonic free states (yellow) for the same

parameters averaged over first 3 nanoseconds. (**e**, **f**) The same as (**c**, **d**) for zero detuning $\hbar\omega - \varepsilon_0 = 0$. (**g**, **h**) The same as (**c**, **d**) for positive detuning $\hbar\omega - \varepsilon_0 = g\sqrt{N} = 30$ μeV.

## 5. Conclusions

In this work, we investigate the interaction of a quantum photon field in a high-Q resonator with a quantum well or a quantum wire of finite size, depending on the detuning of the resonator with respect to the minimal exciton transition frequency. We have found the energies and eigenfunctions of collective excitonic and bi-excitonic states induced by light in a multiatomic system. We demonstrate that the spatial confinement and energy quantization effects arising in 1D and 2D cases provides an important role of the discrete structure of the exciton energy spectrum which should be taken into account. As a result, the time-dependent excitation process is caused by dynamics of population of the collective symmetric modes described in terms of the bosonic ladder operators. However some additional states are found to be populated which are in resonance with several quasienergy levels of the system. The population of these "resonant" is very sensitive to the detuning of field from the minimal exciton transition frequency and leads to significant enhancement of the bi-exciton population. At the same time their role can be suppressed by growing the number of atoms, i. e. by passing to the limit of the infinitely long chain.

It should be also emphasized that the obtained dynamics takes place in case when the dissipative processes are negligibly small. The effects of dephasing lead suppression of the resonant state population and prevents the observation of these states.

It should be noted here that the considered approach assumes two-level atoms, therefore highest excited states of atoms and their ionized states are neglected. This leads to the fact that the biexciton state (as well as states with a large number of excitons) does not undergo Auger decay, in which the states of two excitons pass into the state of an ionized electron-hole pair. Thus, the applicability of considering this model for several excitons is limited by times less than the inverse probability of the Auger interaction. The same limitation applies to other types of interaction of excitons in the system (in particular, exciton-phonon interaction). Taking into account that the typical exciton-phonon interaction time for many systems corresponds to the subpicosecond range, this limitation is the most severe and demand high values of the coupling constant $g$. Nevertheless, these restrictions are not fundamental; the qualitative nature of the interaction with the quantum photon field is similar even if more atomic states are involved in the excitation process.

## 6. Acknowledgment

We acknowledge financial support of the Russian Science Foundation project No 19-42-04105.## 7. References

1. Handbook of Optoelectronics Concepts, Devices, and Techniques, Edited by John P. Dakin and Robert Brown, (CRC Press) 2020 pp. 858
2. J. Homola, Surface Plasmon Resonance Based Sensors, Springer, Berlin, Germany (2006).
3. Belotelov VI, Zvezdin AK. JOSA B. **22**, № 1. pp. 286_292 (2005)
4. Sattler K D (ed) 2010 *Handbook of nanophysics. Nanoelectronics and nanophotonics* (CRC Press)

5. C.F. Bohren, D.R. Huffman Absorption and scattering of light by small particles 1983, Wiley Prof. Edition
6. M. O. Scully and M. S. Zubairy 1997 Quantum Optics (Cambridge: Cambridge University Press)
7. Kira M and Koch St W 2012 *Semiconductor quantum optics* (Cambridge University Press)
8. Hargart F, Roy-Choudhury K, John T, Portalupi S L, Schneider C, Höfling S, Kamp M, Hughes S, Michler P 2016 Probing different regimes of strong field light–matter interaction with semiconductor quantum dots and few cavity photons *New J. Phys*. **18** 123031 doi: 10.1088/1367-2630/aa5198
9. Yu P and Cardona M 2010 *Fundamentals of Semiconductors: Physics and Materials Properties* (Springer)
10. M. Tavis and F. Cummings. Phys. Rev., 170(2), 1968.
11. M. Tavis and F. Cummings. Phys. Rev., 188(2), 1969.
12. B. Garraway. Phil. Trans. R. Soc. A, 369:1137–1155, 2011.
13. A. Retzker, E. Solano, and B. Reznik Phys. Rev. A **75**, 022312, (2007)
14. M. Benedict. Super-radiance: Multiatomic coherent emission. CRC Press, 1996.
15. Sete E A and Eleuch H 2010 Interaction of a quantum well with squeezed light: Quantum-statistical properties *Phys. Rev. A* **82**, 043810 doi: 10.1103/PhysRevA.82.043810
16. Sete E A, Das S, Eleuch H 2011 External-field effect on quantum features of radiation emitted by a quantum well in a microcavity *Phys. Rev. A* **83**, 023822
17. Kasprzak J, Reitzenstein S, Muljarov E A, Kistner C, Schneider C, Strauss M, Höfling S, Forchel A, Langbein W 2010 Up on the Jaynes–Cummings ladder of a quantum-dot/microcavity system *Nature Materials* **9** 304 doi: 10.1038/nmat2717
18. Hopfmann C. Carmele A, Musiał A, Schneider Ch, Kamp M, Höfling S, Knorr A, Reitzenstein St 2017 Transition from Jaynes-Cummings to Autler-Townes ladder in a quantum dot–microcavity system *Phys. Rev. B* **95**, 035302 doi: 10.1103/PhysRevB.95.035302